\documentclass{ieeeaccess}
\usepackage[caption=false]{subfig}
\usepackage[pdftex]{graphicx} 
\usepackage[T1]{fontenc}
\usepackage{cite}

\usepackage{amsmath,amssymb,amsfonts}
\usepackage{algpseudocode, algorithm,algorithmicx}
\usepackage{braket}

\def\BibTeX{{\rm B\kern-.05em{\sc i\kern-.025em b}\kern-.08em
    T\kern-.1667em\lower.7ex\hbox{E}\kern-.125emX}}

\graphicspath{{./figures/}}
\DeclareGraphicsExtensions{.pdf,.jpeg,.png}
\usepackage{color}
\usepackage{soul}

\begin{document}
\history{Date of publication xxxx 00, 0000, date of current version xxxx 00, 0000.}
\doi{10.1109/TQE.2020.DOI}

\title{Multi-User Entanglement Distribution in Quantum Networks Using Multipath Routing}

\author{\uppercase{Evan~Sutcliffe}\authorrefmark{1}, \IEEEmembership{Member, IEEE}, and
\uppercase{Alejandra~Beghelli\authorrefmark{1}} \IEEEmembership{Member, IEEE}}
\address[1]{ Optical Networks Group, Electronic \& Electrical Engineering Department, University College London, London, WC1E 7JE, UK}
\tfootnote{The work of E. Sutcliffe was supported by the Engineering and Physical Sciences Research Council (EPSRC) grant (EP/S021582/1) and (EP/R035342/1). The work of A. Beghelli was supported by EPSRC TRANSNET (EP/R035342/1) and Innovate UK Quantum Data Centre of the Future (10004793) grants
}
\markboth
{Author \headeretal: Preparation of Papers for IEEE Transactions on Quantum Engineering}
{Author \headeretal: Preparation of Papers for IEEE Transactions on Quantum Engineering}

\corresp{Corresponding author: Evan Sutcliffe (evan.sutcliffe.20@ucl.ac.uk).}

\markboth{Journal of \LaTeX\ Class Files,~Vol.~x, No.~x, x~x}%
{}


\begin{abstract}
Quantum networks facilitate numerous applications including secure communication and distributed quantum computation by performing entanglement distribution. For some multi-user quantum applications access to a shared multipartite state is required. We consider the problem of designing protocols for distributing such states, at an increased rate. For this, we propose three protocols that leverage multipath routing to increase the distribution rate for multi-user applications. The protocols are evaluated on quantum networks with NISQ constraints, including limited quantum memories and probabilistic entanglement generation. Simulation results show that the developed protocols achieve an exponential increase in the distribution rate of multipartite states compared to single path routing techniques, with a maximum increase of four orders of magnitude for the cases studied. Further, the relative increase in distribution rate was also found to improve for larger sets of users. When the protocols were tested in scaled-down real-world topologies, it was found that topology had a significant effect on the multipartite state distribution rates achieved by the protocols. Finally, we found that the benefits of multipath routing are maximum for short quantum memory decoherence times and intermediate values of entanglement generation probability. Hence, the protocols developed can benefit NISQ quantum network control and design. 
\end{abstract}

\maketitle

\IEEEpeerreviewmaketitle

\section{Introduction}
\PARstart{A} quantum network is a collection of devices which can exchange quantum information over quantum channels \cite{wehner2018quantum}. This can be achieved by first distributing a shared entangled state between the users that wish to exchange quantum information, and then performing quantum teleportation \cite{cuomo2020towards,bennett1996purification}. Communicating between two users requires a two-qubit (bipartite) entangled state. For multiple users to have access to a shared entanglement, a multi-qubit (multipartite) state must be distributed. Applications which can use shared multipartite states include clock synchronisation \cite{ren2012clock}, distributed quantum sensing \cite{zhang2021distributed}, secret sharing \cite{share,bell2014experimental}, and multi-party Quantum Key Distribution (QKD) \cite{QKDGHZ}. 
A further key motivation for quantum communication is quantum computation, due to the benefits of running quantum algorithms distributed over multiple quantum computers \cite{cacciapuoti2019quantum,cirac1999distributed,alpha_vqe_dist}. In such cases, multipartite states can be used to facilitate multi-qubit operations or for quantum error correction between multiple devices \cite{Zhou2019,nickerson}. \\

Sharing multipartite states between distant users requires the design of multi-user entanglement distribution protocols. Many protocols assume the generation of bipartite entanglement between a central device and each user, which are then transformed into a multipartite state by performing local operations in the central device \cite{bugalho2023distributing, nain2020analysis,netsquid}. If the user is not directly connected to the centre node by a network edge, a long-distance entanglement can be distributed by entanglement swapping, along a pre-computed route of quantum repeaters \cite{repeater1,munro2015inside,briegel1998quantum}. A key drawback to such approaches is that in quantum networks, pre-computed single path routing has a low rate of success, which decreases with the distance between users. This issue is compounded when sharing entanglement between multiple users. A secondary drawback is that the number of quantum memories at the central device can constrain the number of users an entangled state can be distributed between.\\

We propose three multi-user entanglement distribution protocols which can overcome some of the limitations of using single pre-computed paths. The proposed protocols perform routing by dynamically selecting a path, using knowledge of the successfully distributed entanglement states. In designing such protocols we consider current quantum computers, described as \hbox{Noisy Intermediate Scale Quantum (NISQ)} devices due to their limited number of qubits and noisy operations. Therefore, we consider multipartite state distribution protocols for networks constrained by their available quantum resources. Due to their uses in quantum computation and secret-sharing applications \cite{d2005computational, secret}, the protocols developed focus on the distribution of the maximally entangled Greenberger–Horne–Zeilinger (GHZ) states. \\

The remainder of this paper is organised as follows: Section \ref{sec:previous} discusses previous work and highlights the contribution of this paper. Section \ref{sec:model} describes the network model and assumptions whilst Section \ref{sec:statement} gives the problem statement for this paper. The protocols are presented in Section \ref{sec:proposal} and the performance evaluation results are reported in Sections \ref{sec:results1}-\ref{sec:top}. Analytical upper bounds and approximations for the distribution rate are derived in Section \ref{sec:analytical} with concluding remarks in Section \ref{sec:conclusions}.

\section{Previous Work and Contribution} \label{sec:previous}






We classify previous work using two main features: the size of the entangled state to be distributed (bipartite vs. multipartite), and the routing strategy used (single path vs. multipath). For \textit{single path} (SP) routing, Bell pair generation is attempted over the network edges of a unique pre-computed path (or tree for the multipartite case). In contrast, \textit{multipath} (MP) routing attempts Bell pair generation over all network edges and selects the best possible path or tree using only those edges where Bell pairs are present. For ``single path" and ``multipath" we use the term ``path" loosely, as a tree is not a path.

\subsection{Single Path routing for bipartite states}

In a quantum network, an SP routing protocol works by selecting 
 a single path of quantum channels 
which connects two 
users, such that end-to-end distribution rate is maximised \cite{azuma,bauml2020linear, munro2015inside}.
This approach parallels that of shortest-path routing in classical networks. For a network of noisy quantum channels, approaches can be taken to improve the distribution rate and fidelity of the distributed states \cite{cuomo2020towards,li2020efficient}.

\subsection{Single Path routing for multipartite states}
For multipartite state distribution, the routing is necessarily more complex. Multipartite states of $N$ qubits can be distributed to a set of $S$ users, where $ |S| \leq N$. We focus on the case $N = |S|$ with each party receiving a single qubit of the multipartite state.
Some multipartite SP routing protocols extended the concept of bipartite SP routing by pre-calculating paths between the users and a central device \cite{bugalho2023distributing,avis2023analysis, nain2020analysis,netsquid} and then generating a Bell pair between the central device and each of the users. A multipartite state can then be generated from these Bell pairs using only local (qubit) operations and classical communication (LOCC). The route selection can also be performed with secondary parameters such as fidelity or time delay \cite{bugalho2023distributing}. 
By allowing SP routing along a tree of edges connecting the users, a central device is no longer required. However, this approach has an additional classical communication cost \cite{meignant2019distributing,bugalho2023distributing,bauml2020linear}.

\subsection{Multipath routing for bipartite states}

Using multipath routing, a practical entanglement distribution protocol was developed by Pant \textit{et al.} \cite{pant2019routing}, building on results by Pirandola \cite{pirandola2019end} and  Ac{\'\i}n \textit{et al.} \cite{acin2007entanglement}. In the grid topologies studied, the MP protocol achieved a higher end-to-end distribution rate than those that employ SP routing strategies. Additionally, the end-to-end distribution rate did not degrade with distance among users, as long as the Bell pairs between adjacent nodes were generated above a given threshold probability. 
This is a significant improvement on the exponentially decaying \hbox{rate-distance} relationship achieved by SP routing \cite{Pirandola2017}.\\

This distance-independent behaviour of MP routing can be explained in terms of the bond percolation problem. For certain graphs where edges are created probabilistically, a giant connected component (GCC) of $O(|V|)$ nodes emerge when edges are generated above a critical threshold probability $p_{c}$ \cite{molloy_reed_1998,grimmett2013percolation,stauffer2018introduction}, where $|V|$ is the number of network nodes. For grid lattice topologies $p_{c} = 0.5$ and hence percolation is observed for $p>0.5$ In quantum networks, Bell pairs between adjacent nodes can represent the probabilistically generated edges of the bond percolation problem. Therefore, two nodes being in the same connected component means a path of edges exists where every edge holds an entanglement link. By performing BSMs at all nodes along such a path, a long-distance Bell pair can be distributed between these users. When Bell pairs between adjacent nodes are generated above $p_{c}$, the likelihood of a path existing is independent of the distance between the nodes and hence the rate of entanglement distribution is independent of distance.\\

The work of Pant \textit{et al.} \cite{pant2019routing} has since been extended by other authors such as improved protocols for networks of imperfect repeaters, multi-timestep network models, or for sharing 3-qubit GHZ states between two users \cite{laurenza2022rate,patil2021,patil2020,zhao2021redundant}. Some multipath protocols require global knowledge of the distribution of the entangled states over the network \cite{pant2019routing}. 
Other protocols are developed to utilise local knowledge only \cite{patil2020}. This latter approach can reduce the classical communication requirements. 
However, this approach generates large intermediate entangled states, which can reduce the fidelity of the distributed state.\\

\subsection{Multipath routing for multipartite states}
Multipath routing has not been applied to distributing multipartite states between multiple users, except for preliminary results by the authors \cite{sutcliffe2023ofc,sutcliffe2022quantum}.
In \cite{sutcliffe2023ofc} two MP protocols, MP-G and MP-C, for multipartite entanglement distribution are proposed and evaluated in grid topologies. MP-G  extends previous work using a central node by allowing multiple paths to be considered between the central node and each user. The paths are selected considering all edges for which Bell pairs between adjacent nodes are present. MP-C (also discussed in this paper) discards the use of a central node by performing routing using the Steiner tree which connects all the users. Both MP protocols achieved significantly higher multipartite distribution rates than SP approaches. The work in \cite{sutcliffe2023ofc} was extended in \cite{sutcliffe2022quantum} to evaluate the performance of MP-G and MP-C in real-world topologies, where the better performance of MP routing was again confirmed.\\

This paper extends our previous work by:
\begin{itemize}
    \item Proposing the new protocols MP-P and MP-G+.
    \item Describing the proposed protocols in detail by means of pseudo-code and providing expressions for their routing computational complexity and classical communication complexity.
    \item Evaluating the performance of the protocols under new scenarios
     \item Extending the results in \cite{sutcliffe2023ofc} by evaluating the performance of the new MP-P protocol in mesh topologies.
    \item Providing an analytical approximation of the distribution rate for the best performing MP protocols.
\end{itemize}

Overall, the contribution of this paper 
consists of presenting consolidated results that address for the first time the problem of multi-user entanglement distribution using multipath routing.

\section{Network model and assumptions} \label{sec:model}

\subsection{Quantum network model}\label{sec:modelmodel}

\Figure[ht!][width=0.9\textwidth]{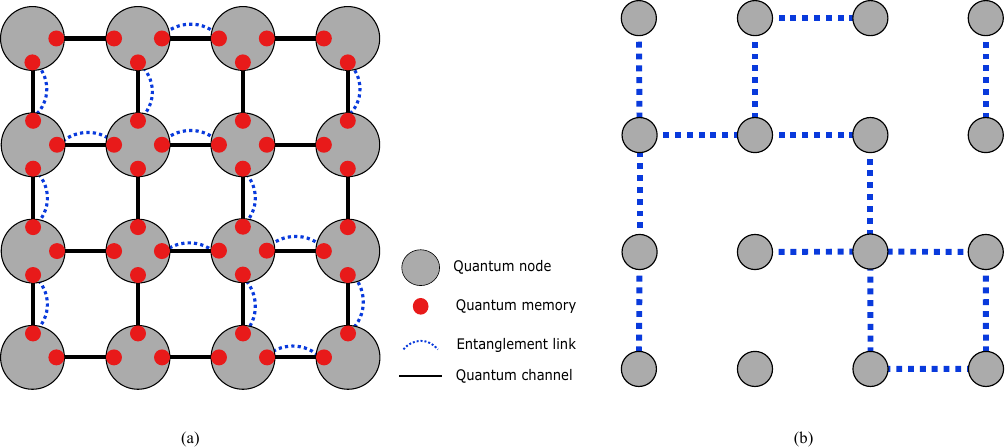}
{a) Diagram of a quantum network defined by a graph $G$ in a $4 \times 4$ grid topology. Also shown is a random set of entanglement links shared between adjacent nodes. b) Subgraph $G' =(V,E')$ with edges from $G$ with $\omega(e) = 1$.\label{fig:network}}

A quantum network can be represented as a graph \hbox{$G = (V, E)$}, with a set of nodes $V$ and edges $E$. An example $4 \times 4$ grid topology is shown in \hbox{Fig. \ref{fig:network}a)}. \\

\textbf{Edges} represent quantum channels, over which \textit{entanglement links} can be generated. An entanglement link is a maximally entangled two-qubit state, shared between adjacent nodes in a quantum network. We assume entanglement links are distributed in the form of a \hbox{$\ket{\phi^+} = \frac{1}{\sqrt{2}} (\ket{00} +\ket{11})$} state. For clarity, we henceforth refer to the \hbox{$\ket{\phi^+}$} state as an \textit{entanglement link} when shared between adjacent nodes \hbox{(as in Fig. \ref{fig:network})}, and as a \textit{Bell pair} when shared between distant users by entanglement swapping.\\

Distributing entanglement links over a noisy quantum channel is lossy and hence probabilistic. The probability $p_e$, of successfully generating an entanglement link over an edge $e \in E$ can be modelled as:
\begin{equation} \label{eq:1}
 p_e = p_{\text{op}} (1-p_{\text{loss}})
\end{equation} 
where $p_{\text{op}}$ denotes the probability of imperfect node operations in entanglement link generation and $p_{\text{loss}}$ the probability of qubit loss in the channel \cite{munro2015inside}. If we assume photonic qubits with channels of optical fibre, then for a channel of length $L$ km with attenuation $0.2$ dB/km, this loss can be expressed as $p_{\text{loss}}=1-10^{-0.2 L/10}$.
The operation probability $p_{\text{op}}$ represents a lumped probability of generating an entanglement link for two back-to-back devices (e.g. at $L = 0$km), thus excluding photon loss in the fibre from $p_{\text{op}}$. Factors that can affect $p_{\text{op}}$ include failure in photon generation, imperfect qubit-photon entanglement or photon frequency conversion \cite{munro2015inside,freq_convert}. \\

\textbf{Nodes} represent devices able to store qubits in quantum memories and perform LOCC. We assume all nodes have equal capabilities and can perform any function (e.g. as a user, repeater or centre node). Thus, any nodes in the network can request to share a multipartite state. The capabilities are as follows:
\begin{itemize}
    \item All nodes have a single quantum memory per edge, allocated for communication purposes. 
    \item When an entanglement link is successfully generated over a network edge, entangled qubits are stored in specified quantum memories at the adjacent nodes connected to that edge. While the quantum memories are occupied by an entanglement link, they cannot be reused for a new entanglement link generation process.
    \item All quantum memories have identical decoherence times. The quantum memories are modelled using a cut-off decoherence of time $T_c$. That means that before $T_c$, a qubit is stored in the quantum memory with perfect fidelity. After $T_c$ the qubit is assumed to have undergone decoherence and is discarded \cite{azuma}.
    \item All LOCC operations are error-free, e.g. all nodes can perform error-free local operations between qubits stored in any quantum memory. Error-free classical communication is enabled by a parallel classical network.\\
\end{itemize}

The main LOCC operations used for distributing multipartite states are entanglement swapping and entanglement fusion. Entanglement swapping allows for the distribution of long-distance entanglement along a path of quantum repeaters. As we assume the devices can freely select any two qubits when performing entanglement swapping, these nodes could also be defined as quantum switches\footnote{This definition of a quantum switch should not be confused with a different definition, in which a quantum switch refers to using a qubit to control the operation order of a circuit \cite{switch1}.} \cite{lee2022quantum,patil_plus}. 
A Bell pair can be shared between distant nodes by entanglement swapping \cite{repeater1,munro2015inside,briegel1998quantum}. By performing a Bell State Measurement (BSM) on the qubits of entanglement links at each intermediate node along the path, a long-distance Bell pair is distributed, such as shown in Fig. \ref{fig:swap}.\\ 

\Figure[h!t][width=80mm]{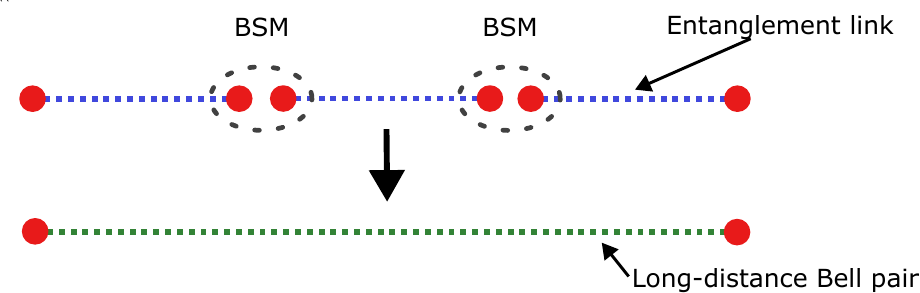}{Entanglement swapping is performed to distribute a long-distance Bell pair from a path of entanglement links. This is achieved by performing BSMs on the qubits at intermediate nodes. Entanglement links and Bell pairs are both $\ket{\phi^+}$ states \label{fig:swap}}

\begin{figure*}[ht]%
    \centering
    \subfloat[\centering]
    {{\includegraphics[height=3.3cm]{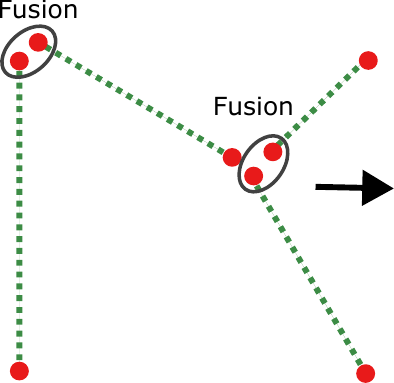}}}
    \qquad
    \subfloat[\centering]{{\includegraphics[height=2.966cm]{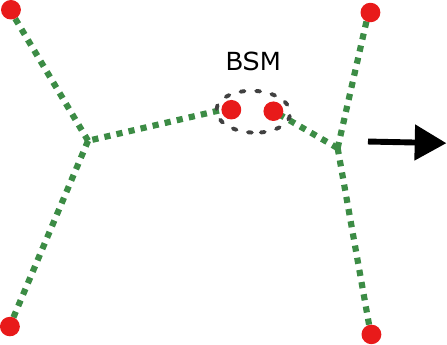}}}
    \qquad
    \subfloat[\centering]{{\includegraphics[height=2.945cm]{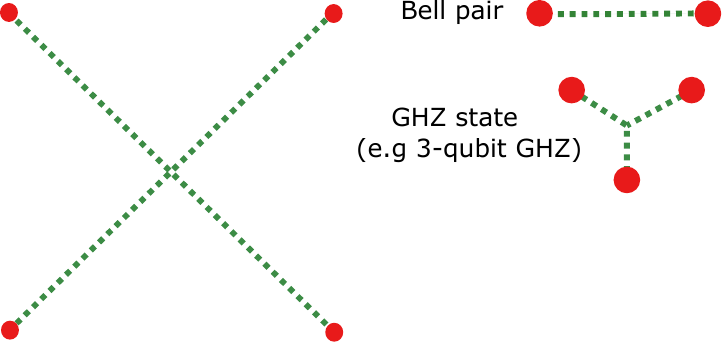}}}
    \qquad
    \caption{Example of the distribution of a  $\ket{\text{GHZ}_{4}}$ state from four Bell pairs. GHZ states are shown using a star symbology, which is not equivalent to the graphical notation of a graph state. a) Two entanglement fusion operations are performed, combining four Bell pairs into two $\ket{\text{GHZ}_{3}}$ states. b) Two $\ket{\text{GHZ}_{3}}$ states are then combined by a BSM to generate a $\ket{\text{GHZ}_{4}}$ between the desired qubits.\label{fig:fusion} }%
\end{figure*}

A $N$-qubit GHZ state can be generated by entanglement fusion using at least $N-1$ Bell pairs. The \hbox{$N$-qubit GHZ} state is given as \hbox{$\ket{\text{GHZ}_{N}} = \frac{1}{\sqrt{2}}(\ket{0}^{\otimes N}  + \ket{1}^{\otimes N})$}. The $\ket{\phi^+}$ state is equivalent to a $\ket{\text{GHZ}_{2}}$ state. The entanglement fusion operation combines two GHZ states of qubit sizes $n_1$ and $n_2$ and entangles them, to generate a single state of size $n_1+n_2-1$. This operation is executed at a node by entangling a qubit from each state, by measuring one of the qubits and then performing corrections depending on the measurement outcome. These deterministic operations can be performed iteratively to generate large multipartite states \cite{de_Bone_2020,meignant2019distributing}. Similarly, a BSM can combine two GHZ states into a single state of size $n_1+n_2-2$ \cite{avis2023analysis}. \hbox{Fig. \ref{fig:fusion}} shows the generation of a four-qubit GHZ state from four Bell pairs. By generating GHZ states from multiple Bell pairs, multipartite states can be shared across a quantum network without requiring all qubits to be successfully transmitted along a separate point-to-point connection for each user.\\



 \subsection{Quantum Network Operation Assumptions}\label{sec:modeloperation}
 

\textbf{Link-state information}. The binary variable $\omega(e) \in \{0,1\}$ is a state which represents if an entanglement link is present $\omega(e) = 1$, or absent $\omega(e) = 0$ over an edge $e$. The subgraph $G' =(V,E')$, as shown in \hbox{Fig. \ref{fig:network}b}, can represent the global link-state, of $\omega(e)$ for all $e\in E$ over a network. The edges $E'$ are the subset of $E$ for which \hbox{$\omega(e) = 1$}. We assume that global link-state information $G'$ is available for network operations. 
The knowledge of the link-state is made possible by heralded entanglement distribution, where success is flagged by a classical signal \cite{heralded}. Collating the global link-state information will have an associated classical communication and time delay cost. Evaluating such costs is out of the scope of this work.\\


\textbf{Time-slotted operation}. We assume a discrete-time network operation model.
Each timeslot lasts for time $T_{\text{slot}}$, and thus an entanglement link can be stored (i.e. $\omega(e) = 1$) over an over edge for up to $Q_c$ timeslots \hbox{($Q_{c} = \lfloor T_{c}/T_{\text{slot}} \rfloor$)} from generation \cite{li2020efficient}. After an entanglement link has been stored for over $Q_c$ timeslots, it is discarded ($\omega(e) = 0$). This means that $\omega(e)$ varies over multiple timeslots.
This temporal variation is described by the sequence $\Omega_T=(\omega(e)_1,\omega(e)_2, ... \omega(e)_T)$, where $\omega(e)_t$ is the state of $\omega(e)$ in timeslot $t$ and $T$ is the network operation period, measured in number of timeslots. 
If a \hbox{$N$-qubit GHZ} state is not successfully established in one timeslot, the protocol reattempts distribution in future timeslots. In these future timeslots, only entanglement links that are still present (i.e. within $Q_c$ timeslots since generation) plus newly generated entanglement links can be used to find a routing solution $R$.\\ 


\section{Problem Statement} \label{sec:statement}
Consider a quantum network represented by the graph \hbox{$G = (V, E)$} with a subset of vertices $S\in V$, requesting an N-qubit GHZ state to be shared among them. 
In this work, the problem of \textbf{multipartite entanglement distribution} consists of generating a $\ket{\text{GHZ}_{N}}$ state between users $S$ with $|S|>2$ and $N=|S|$, such that the rate at which the GHZ state is generated per time slot is maximised. We call this the \textit{distribution rate} and denote it by DR.

To generate a $\ket{\text{GHZ}_{N}}$ state, a routing solution $R$ must be found. The routing solution represents a set of entanglement links which can be combined by LOCC operations to generate the required GHZ state. A \textbf{multipartite entanglement distribution protocol} specifies the set of rules for finding a solution to the problem described above.

We assess the performance of the protocols developed in terms of the DR, defined as the average number of GHZ states distributed per timeslot $\text{DR} = \frac{\#{\text{GHZ}}}{T_{\text{slot}}}$. The rate that entangled states are distributed upper bounds the quantum information transfer in a quantum network \cite{Pirandola2017}. This metric does not take into account the size of the GHZ state distributed. Further, the DR should not be confused with the rate at which entanglement links are distributed between adjacent nodes.\\ 

\section{Multipartite distribution protocols} \label{sec:proposal}

\subsection{Benchmark Solution (SP protocol)} \label{sec:SP}
For comparison purposes, a generalised version of a multipartite single path protocol is used as a benchmark, denoted as the SP protocol. The SP protocol utilises a central node and only attempts entanglement link generation along pre-calculated shortest paths from a central node to each user \cite{bugalho2023distributing,avis2023analysis}. These paths together describe the routing solution $R$.\\ 

The operations performed by the SP protocol are described in Algorithm \ref{alg:sp}. 
Initially, the centre node $v_c$ is selected (line 2) using an exhaustive strategy: for each candidate centre node $v \in V$, with a nodal degree ($deg(v)$) greater than or equal to $|S|$, a routing solution $R_v$ is found using a max-flow routing algorithm \cite{megiddo1974optimal}. Next, the centre node $v_c$ is selected such that ${DR}_{\text{SP}}(v_c) \geq DR_{\text{SP}}(v) \; \forall  \; v \in V$ with $deg(v) \ge |S|$. For the SP protocol, the value of $DR_{\text{SP}}(v)$ can be found directly from the routing solution: 
\begin{equation} \label{eq:prod}
  \text{DR}_{\text{SP}}(v) = \prod_{e \in R_v} p_e
\end{equation}

If no valid routing solution $R_v$ can be found for the candidate node $v$ then $DR_{SP}(v) = 0 $. A valid routing consists of an edge-disjoint path between each user and the centre node. The paths must be edge-disjoint due to the assumption that nodes can only store a single entanglement link per edge and because entanglement links are consumed by entanglement swapping. As there must be $|S|$ edge-disjoint paths between $v_c$ and each user in $S$, the centre node must have a nodal degree greater than or equal to $|S|$.\\ 

The selection of the centre node $v_c$ is performed before multipartite state generation is attempted and remains fixed throughout the operation of the protocol. In grid topologies with uniform $p_e$, the centre node selection is reduced to selecting the centroid of the users. The routing solution computed for the selected central node is stored as $R$ (line 3).\\
 
\begin{algorithm}
    \caption{SP protocol}
    \begin{algorithmic}[1]
    \Function{SP}{$G, S$}
    \State $v_c$ = selectCentreNode(G,S) 
    \State $R =$ getShortestPaths($G,v_c,S$) 
    \State hasGHZ $= \mathbf{False}$ 
    \While{$\textbf{not}$ HasGHZ}
        \State  SimulateEntanglementLinks(G) 
        \State $G'$ =  updateLinkSubgraph(G)    
        \State $S'$= $S - $\{ hasSharedBellPair($G, v_c, S$) \} 
        \For{$s \in S'$}
            \If{$R[s] \in G'$} 
                \State entanglementSwapping($G , R[s] , v_c, s$)
                \State $G'$ =  updateLinkSubgraph(G)  
            \EndIf
        \EndFor
        \If{hasSharedBellPair$(G, v_c , S) == S$} 
            \State entanglementFusion($G , v_c , S$) 
            \State hasGHZ $= \mathbf{True}$
        \EndIf
    \EndWhile 
    \EndFunction
    \end{algorithmic}\label{alg:sp}
\end{algorithm}

Next, the protocol runs for multiple timeslots, terminating once a GHZ state is generated (lines 5-19). At the start of each timeslot, the entanglement link generation is attempted over all edges in $R$, and qubit decoherence of the network model is simulated (line 6), meaning any entanglement links older than $Q_c$ timeslots are discarded. The state of $G'$ is then updated (line 7). The protocol operates to generate a Bell pair shared between the centre node and each user. To do so the protocol first obtains the subset of users $S'$ which do not currently hold a Bell pair shared with the centre node (line 8). For each of these users $s \in S'$, the pre-computed path  $R[s]$ between $s$ and the centre node is checked to assess if all edges in the path $R[s]$ hold an entanglement link (line 10). If so, a Bell pair is generated by performing entanglement swapping along the path (line 11), and $G'$ is updated (line 12). 
When all users share a Bell pair with the centre node (line 15), the GHZ state is generated by performing entanglement fusion between the qubits of the Bell pairs held centre node (line 16).\\ 

\subsection{Proposed Multipath Protocols}
We propose three protocols, which are novel variants of multipath routing applied to distributing a multipartite state between multiple users. 
The proposed multipath (MP) protocols are the Greedy Plus \hbox{(MP-G+)}, Cooperative \hbox{(MP-C)}, and Packing \hbox{(MP-P)} protocols. 

In each timeslot, the MP protocols perform three distinct operations:
\begin{itemize}
    \item Entanglement link generation: nodes attempt to generate entanglement links over all the edges in $G$. 
    \item Multipath routing: the protocols attempt to compute a \textit{routing solution} using the global link-state information represented as the sub-graph $G'$. Unlike SP, where the routing solution is made of edges in G, the routing solution of MP protocols only consists of edges in $G'$, which are known to hold entanglement links.
    \item GHZ state generation: if a routing solution was found, a $N$-qubit GHZ state is generated from the selected entanglement links. This is done such that the qubits of the GHZ state are shared among the users $S$.
\end{itemize}

\subsubsection{Multipath Greedy (MP-G+)}

\begin{figure*}[htbp]%
    \centering
    \includegraphics[width=\textwidth]{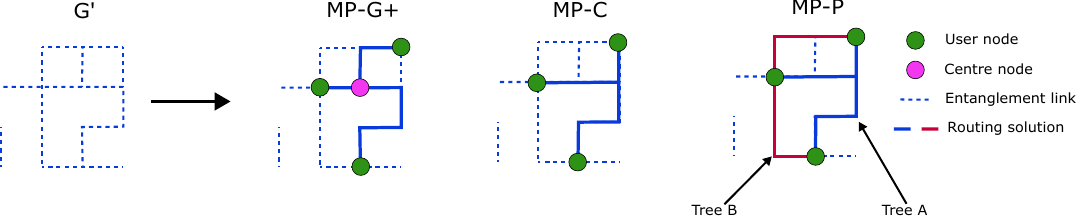}
    \caption{Routing solutions for the MP-G+, MP-C and MP-P protocols for three users in an example state of $G'$. The MP-G+ protocol finds the minimum distance edge-disjoint paths in $G'$ to connect the users to a pre-selected centre node. For this example, this requires 7 entanglement links. The MP-C and MP-P both use a routing solution consisting of the Steiner tree (using 6 links). However, for the MP-P protocol, multiple GHZ states can be generated if multiple Steiner trees can be found in $G'$. The example shows a situation where two  $\ket{\text{GHZ}_{3}}$ states can be generated (Tree A \& B).}
    \label{fig:protocols}%
\end{figure*}

The MP-G+ protocol extends the ideas of the SP protocol described in Section {\ref{sec:SP}} by allowing multipath routing. Similarly to the SP protocol, a valid routing solution consists of an edge-disjoint path between the centre node and each user. However, whereas the SP protocol uses a single pre-computed set of paths in which entanglement link generation is reattempted each timeslot, the MP-G+ protocol can select paths from any edges in $G'$. By routing paths over $G'$, the routing solution only uses edges which hold entanglement links. For the multipath protocols, a routing solution is valid if a route can be found from edges that hold entanglement links. Each path of entanglement links can be used to generate a Bell pair. Once a Bell pair has been shared with each user, a GHZ state is then generated by performing entanglement fusion at the centre node.\\

The MP-G+ protocol is described in Algorithm \ref{alg:mpg+}. First, a centre node is selected using the same sub-procedure as for the SP protocol (line 2). This selection maximises the DR for the SP protocol, therefore is considered a suitable centre node for the MP-G+ protocol.\\

Next, the MP-G+ protocol executes over multiple timeslots to generate a GHZ state (lines 4-17). At the start of a timeslot, the entanglement link generation and qubit decoherence are simulated (line 5) and the state of $G'$ (line 6) is updated. Then, the protocol identifies the subset $S'$, made of users that do not currently share a Bell pair with the centre node (line 7).  A routing solution of valid paths is then found between the centre node and as many users in $S'$ as possible while using the fewest number of entanglement links (line 8). To do so, the set of edge-disjoint paths between a single-source (centre node) and multi-sinks (users in $S'$), where each sink can only utilise a single path, are computed for each timeslot using a max-flow approach algorithm \cite{karp,megiddo1974optimal}.  Routing using this approach is an improvement to the initial Multipath Greedy (MP-G) protocol proposed by the authors in \cite{sutcliffe2022quantum}, where a path between the centre node and each user was found iteratively as the shortest path in $G'$. Notice that a valid path might not be found for every user in $S'$ in the same timeslot. In practice, the MP-G+ was found to achieve only a small improvement in the DR compared to the original MP-G protocol \cite{sutcliffe2022quantum}.\\

For each path stored in $R$, a Bell pair is shared between the centre node and the specific user by entanglement swapping (line 10) and the link-state information in $G'$ is updated (line 11). If all users in $S$ share a Bell pair with the central node, then a GHZ state is finally generated (lines 13-14).\\
 
An example routing solution of the MP-G+ protocol is shown in \hbox{Fig. \ref{fig:protocols}}, where it can be seen that the shortest paths in $G'$ are not necessarily the shortest paths in the underlying network topology $G$.\\

\begin{algorithm}
    \caption{MP-G+ protocol}\label{alg:mpg+}
    \begin{algorithmic}[1]
    \Function{MP-G+}{$G, S$}
    \State $v_c$ = selectCentreNode(G,S) 
    \State hasGHZ $= \mathbf{False}$ 
    \While{$\textbf{not}$ HasGHZ} \label{agline:l1}
        \State  simulateEntanglementLinks(G) \label{agline:2}
        \State $G'$ =  updateLinkSubgraph(G) \label{agline:1}
        \State $S'$= $S -$ \{ hasSharedBellPair($G, v_c, S$) \} \label{agline:5}
        \State $R =$ getShortestPaths($G',v_c,S'$) \label{agline:6}
        
        \For{$path \in R$}
            \State entanglementSwapping($G , path , v_c, s$)
            \State $G'$ =  updateLinkSubgraph($G$)
        \EndFor
        \If{hasSharedBellPair$(G, v_c , S) == S$} 
            \State entanglementFusion($G , v_c , S$) \label{agline:3}
            \State hasGHZ $= \mathbf{True}$
        \EndIf \label{agline:l2}
    \EndWhile 
    \EndFunction
    \end{algorithmic}
\end{algorithm}

\subsubsection{Multipath Cooperative (MP-C)}

The MP-C protocol is a multipartite entanglement distribution protocol, which relaxes the constraint of requiring a central node. Instead, a GHZ state can be generated from a tree of entanglement links, which connects the users in $S$. By using a Steiner tree as the routing solution $R$ a GHZ state can be distributed using the fewest number of entanglement links possible. \hbox{Fig. \ref{fig:protocols}} illustrates the operation of the MP-C protocol where users are connected by a Steiner tree of entanglement links.\\

The MP-C protocol is described in Algorithm \ref{alg:mpc}. The protocol runs for multiple timeslots until a GHZ state is distributed (lines 3-12). At the start of each timeslot, entanglement link generation and the qubit decoherence are simulated (line 4) and G' is updated (line 5). 
The protocol then checks if all users are in the same connected component in $G'$ (line 6). 
This is a sufficient condition for the existence of a connecting tree of entanglement links between them. Therefore, the routing solution is found as the Steiner tree in $G'$ that connects $S$ (line 7). A GHZ state is generated from the entanglement links along the routing solution by entanglement swapping and entanglement fusion operations (lines 8-9). First, by performing entanglement swapping, the entanglement links along $R$ are converted into long-distance Bell pairs which are shared between the users and nodes in the Steiner tree that have a nodal degree greater than two. The GHZ state is generated by performing entanglement fusion operations at all nodes in the Steiner tree which hold multiple qubits (line 9).\\


By routing using a Steiner tree, the MP-C protocol does not require a central node and the routing solution will on average require fewer entanglement links. However, this means that entanglement fusion operations might be required at multiple nodes when generating a GHZ state. The LOCC operations required are therefore more complex compared to the MP-G+ and SP protocols, where entanglement fusion is performed only at the centre node. Additionally, a potential drawback to MP-C is that without a fixed centre node, the protocol must wait until all users can be connected by a Steiner tree in the same timeslot. This contrasts with the MP-G+ and SP protocols, in which the generation of a Bell pair between each $v_c$-user pair can be performed as soon as a path exists, with the GHZ state being generated once all users share a Bell pair with the centre node.


\begin{algorithm}
    \caption{MP-C protocol}\label{alg:mpc}
    \begin{algorithmic}[1]
    \Function{MP-C}{$G, S$}
    \State HasGHZ $= \mathbf{False}$
    
    \While{$\textbf{not}$ HasGHZ}
        \State  SimulateEntanglementLinks(G)
        \State $G'$ =  updateLinkSubgraph(G)
        \If {hasConnectingTree($G' , S$)}
            \State $R = $minimumSteinerTree($G' , S$) 
            \State EntanglementSwapping($G, R , S$) %
            \State EntanglementFusion($G, R , S$) 
            \State HasGHZ $= \mathbf{True}$
        \EndIf
    \EndWhile
    \EndFunction
    \end{algorithmic}
\end{algorithm}

\subsubsection{Multipath Packing (MP-P)}

The MP-G+ and MP-C protocols attempt to distribute a single GHZ state per timeslot. If there are multiple edge-disjoint trees in $G'$, multiple GHZ states can be generated between the same set of users. \hbox{Fig. \ref{fig:protocols}} shows an example where the MP-P protocol can generate two GHZ states for a single instance of $G'$. This can improve the multipartite distribution rate or benefit applications that require multiple copies of a multipartite state. These include QKD \cite{GHZQKD}, or entanglement distillation, when multiple copies of a state can be combined to improve the average fidelity of the output state \cite{bennett1996purification,krastanov2021heterogeneous,de_Bone_2020}.\\

The MP-P protocol is an improvement of the MP-C protocol that exploits the existence of multiple trees to increase the multipartite distribution rate. Thus, instead of terminating after generating a single GHZ state, the GHZ generation operations (lines 6-12) are repeated until a connecting tree can no longer be found. 
The protocol's name derives from the tree-packing problem, for finding the maximum number of Steiner trees in a graph. \\

\begin{algorithm}
    \caption{MP-P protocol}\label{alg:mpp}
    \begin{algorithmic}[1]
    \Function{MP-P}{$G, S$}
    \State HasGHZ $= \mathbf{False}$
    
    \While{$\textbf{not}$ HasGHZ}
        \State  SimulateEntanglementLinks(G)
        \State $G'$ =  updateLinkSubgraph(G)
        \While {hasConnectingTree($G',S$)}
            \State $R$ = minimumSteinerTree($G',S$) 
            \State EntanglementSwapping($G,R,S$) 
            \State EntanglementFusion($G,R,S$)
            \State $G'$ =  updateLinkSubgraph(G)
            \State HasGHZ $= \mathbf{True}$
        \EndWhile
    \EndWhile
    \EndFunction
    \end{algorithmic}
\end{algorithm}

\subsection{Protocol comparison}

A comparison of the main features of the proposed MP protocols and the SP protocol is shown in Table \ref{tab:comp}. The number of nodes and edges in $G$ is given by $|V|$ and $|E|$ respectively. Similarly, $|R|$ gives the number of edges in the routing solution $R$. In terms of scalability, the MP-G+ and SP protocols described require a centre node to share a Bell pair with each user. Hence the size of the GHZ state that can be distributed is limited by the number of quantum memories at the centre node (equal to the nodal degree of the centre node). Therefore, these protocols do not freely scale with the number of users. \\

In terms of computational complexity, we consider the classical computation that must be performed per timeslot. As the SP protocol uses a pre-computed path, the only per-timeslot operation required is verifying that all edges in $R$ hold an entanglement link. In contrast, the multipath protocols attempt routing operations every timeslot. The computational complexity is therefore dominated by the performance of these sub-operations such as the Edmonds-Karp algorithm to calculate the multiple paths for the MP-G+ protocol ($O(|V|^{2}|E|)$) \cite{karp,karp_speedup} or Mehlhorn's approximate Steiner tree algorithm ($O(|E|+|V|log|V|)$) \cite{mehlhorn1988faster}. Finally, in terms of classical communication complexity (number of messages exchanged), the multipath protocols require additional classical communication compared to shortest path protocols, as the state of $G'$ must be obtained.

\begin{table}[ht!] 
    \caption{Comparative summary of multipartite routing protocols.}
    \label{tab:comp}
    \centering
         \begin{tabular}{|p{0.9cm}|p{1cm}|p{3cm}|p{1.8cm}|}
     \hline
    \textbf{Protocol}&\textbf{Free scaling with users}&\textbf{Routing computational complexity (per timeslot)}&\textbf{Classical communication complexity}\\
     \hline
    SP           & No   & $O(|R|)$                    & $O(|R|)$ \\
    MP-G+     & No   & $O(|V|^{2}|E|)$   & $O(|E|)$ \\
    MP-C      & Yes  & $O(|E|+|V|log|V|)$     & $O(|E|)$ \\
    MP-P     & Yes   & $O((|E|+|V|log|V|)|V|)$ & $O(|E|)$ \\ \hline
    \end{tabular}
\end{table}

\section{Performance evaluation: protocols} \label{sec:results1}
The protocols were evaluated using a Monte Carlo simulation run on the quantum network model described in Section \ref{sec:model}. We compare the protocols in terms of the distribution rate. \\

Throughout Section \ref{sec:results1} we evaluated the protocols on a baseline scenario, varying one parameter at a time (i.e. $p_e$, the distance between users, numbers of users, and decoherence time). This baseline scenario was defined as a square grid topology of size $M \times M$  with $M=6$ and $|S|=4$ randomly located users in set $S \in V$. The entanglement link generation probability was uniform for all edges (i.e. $p_e=p$) and fixed at $p=0.75$. The quantum memory decoherence was assumed to be sufficient to only store entanglement for a single timeslot \hbox{(i.e. $Q_c=1$)}. When plotting the data, unless otherwise stated, each data point in figures \hbox{Fig. \ref{fig:1}-\ref{fig:simulated}} represents the DR achieved by the protocols, averaged over $500$ random user locations in the network. Each protocol was executed until a GHZ state was generated, or terminated after $t=5000$ timeslots. If more than 5\% of the protocol runs terminated without generating a GHZ state, the datapoint was not plotted.

\subsection{Effect of entanglement link generation on multipartite state distribution rate} \label{sec:DRp}

\begin{figure}[htbp]%
    \centering
    \includegraphics[width=\linewidth]{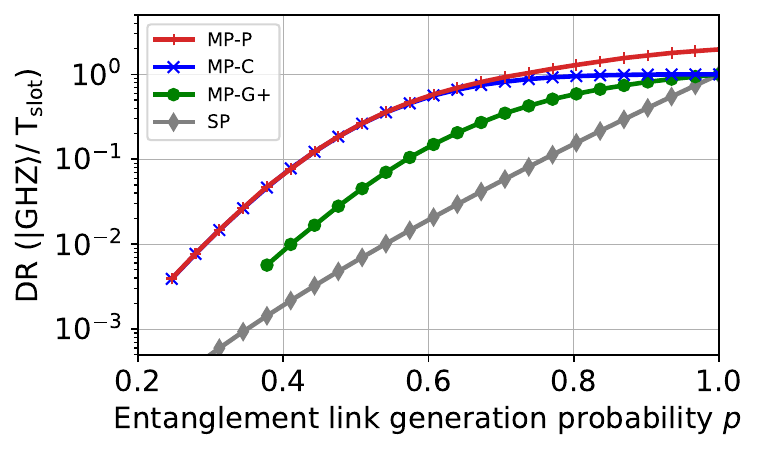}
    \caption{Distribution rate (DR) of  $\ket{\text{GHZ}_4}$ states against entanglement link generation probability $p$ in a $6\times6$ grid with $|S|=4$ randomly located users ($M=6$, 
 $|S| = 4$, $p \in [0,1]$, $Q_c=1$)}
    \label{fig:1}
\end{figure}

Fig. \ref{fig:1} shows the DR of the proposed multipartite protocols as a function of $p$, with the performance of the shortest path (SP) protocol also shown for comparison. It can be observed that all three of the proposed multipartite protocols achieved a higher DR than the SP protocol. The simulation results show the MP-P and MP-C protocols achieved a DR approximately 38 times higher than the SP protocol at $p=0.48$. The more flexible multipath routing means GHZ states can be generated for more of the possible instances of $G'$, e.g., different distributions of successful entanglement links in $G'$. We observe that the MP-P and MP-C protocols also outperform the MP-G+ protocol. This can be justified by considering the routing requirements of the protocols. The MP-P and MP-C protocols can use any Steiner tree of edges in $G'$ for routing, whereas the MP-G+ protocol requires a separate edge-disjoint path between each user and the centre node. As a result, the MP-P and MP-C protocols can generate GHZ state for more possible instances of $G'$. Additionally, all routing solutions that can be used by the MP-G+ protocol can also be used by the MP-P and MP-C protocols. Hence, these protocols will always achieve a DR greater than or equal to the MP-G+ protocol. \\

Finally, we observed that the MP-P protocol outperforms the MP-C protocol for $p \gtrapprox 0.7$, where the MP-P protocol achieved a $\text{DR}>1$. This occurs when on average multiple GHZ states are distributed per timeslot. This condition is met when there are multiple edge-disjoint Steiner trees connecting users in $G'$. As multiple trees are unlikely to exist below the percolation threshold, the MP-P performs comparably to the MP-C protocol for $p<0.5$. 
As the average DR of the SP protocol has an analytical solution for $Q_c=1$, the Monte Carlo simulation was not required to calculate the DR. Instead, the DR was found using (\ref{eq:prod}) for a given routing solution $R$.\\

\begin{figure}[h!t]%
    \centering
    \includegraphics[width=\linewidth]{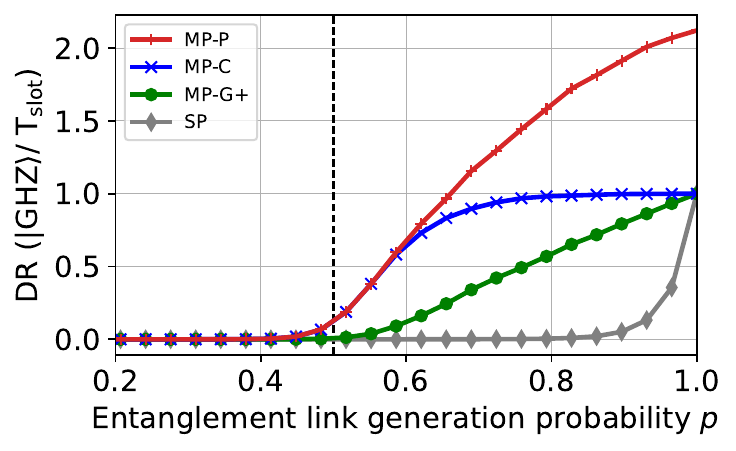}
    \caption{Distribution rate of  $\ket{\text{GHZ}_4}$ states against entanglement link generation probability $p$ in a $18\times18$ grid with $|S|=4$ randomly located users ($M=18$, 
 $|S| = 4$, $p \in [0,1]$, $Q_c=1$)}
    \label{fig:abc}
\end{figure}

To better visualise the phase transition observed on systems exhibiting percolation, Fig. \ref{fig:abc} replicates the results of Fig. \ref{fig:1} but in a larger $18 \times 18$ grid topology. Further, the DR is shown on a linear scale, to better show the phase transition of DR against $p$. However, this approach does not allow for the visualisation of DR for values spanning different orders of magnitude. Figure \ref{fig:abc} shows that the DR of the MP-P and MP-C protocols increase rapidly once percolation is observed ($p>0.5$). For these protocols, the condition for successful routing is equivalent to all users being in the same connected component, which occurs with a high probability once a GCC exists. Because of the extra requirement of edge-disjoint paths, the phase transition is not as clear for the MP-G+ protocol.

\subsection{Distance-independent multipartite state distribution rate} \label{sec:distant}
A key benefit of multipath routing shown in the literature is the ability to distribute entangled states, at a rate independent of the distance between the two users \cite{acin2007entanglement, pant2019routing}. We show that the developed MP protocols also achieve this result for multipartite states shared between multiple users. 
To demonstrate this we simulate the protocols of grid networks of increasing size ($M \times M$ nodes), where the four corners nodes were selected as the users. The selection of corner nodes represents a worst-case scenario for the MP protocols. \\

\begin{figure}[ht]
    \centering
    {\includegraphics[width=\linewidth]{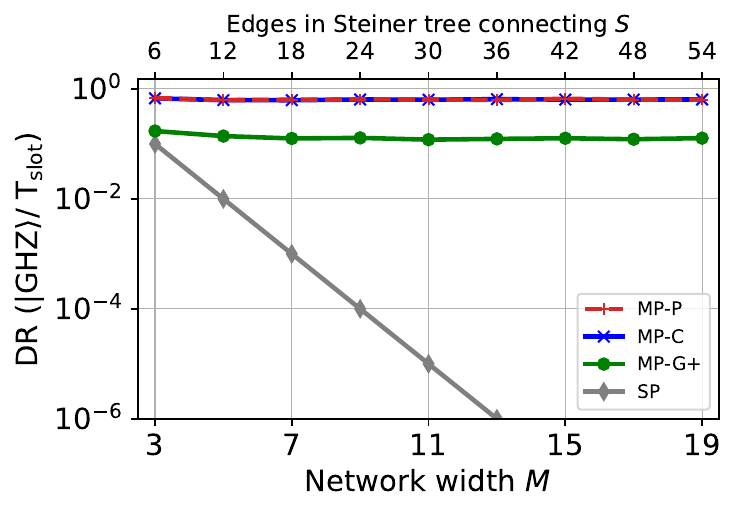}}
    \caption{Distribution rate of $\ket{\text{GHZ}_4}$ states  between the four corner nodes against network grid size $M$ (lower \hbox{x-axis}) and number of edges in the Steiner tree connecting the users (upper \hbox{x-axis}) ($M \in [3,19]$, $|S| = 4$, $p=0.75$, $Q_c=1$)} \label{fig:a} 
\end{figure}

Fig. \ref{fig:a} shows the DR achieved, plotted against the width $M$ of the grid topologies (lower x-axis). We define the number of edges in this Steiner tree in $G$ as the distance between multiple users. For \hbox{Fig. \ref{fig:a}} this Steiner tree consists of $3 \times (M-1)$ edges (value shown on the upper \hbox{x-axis}) and is a lower bound to the number of edges of a routing solution in $G'$.
It can be seen that the MP protocols maintain a DR which is constant with distance. In contrast, the DR achieved by the SP protocol decreases exponentially with distance. Consistent with \hbox{Fig. \ref{fig:1}}, the MP-P and MP-C protocols achieved higher DRs than the MP-G+ protocol.\\

As the entanglement link generation probability is above the percolation threshold $p_{c}=0.5$ of the given topology, the likelihood of all users being in the same connected component is the same regardless of the distance between the users. This is a sufficient condition for a routing solution to exist for the MP-P and MP-C protocols and hence distribution rate is independent of the distance between the users. These results show that the developed protocols can achieve distance-independent DR, even for multiple users sharing multipartite states. We use the metric \textit{speedup} to quantify the relative performance of a MP protocol, compared to the benchmark SP protocol. The speedup is defined by the ratio of the DR achieved by a protocol, in comparison to the benchmark SP protocol under identical network conditions:
\begin{equation}
    \text{speedup} = \frac{DR_{\text{protocol}}} {DR_{\text{SP}}} 
\end{equation} 
The speedup observed by the MP protocols was found to be of order \hbox{$O((1/p)^{|S|})$ ($p_{c} < p \leq1$, $|S| \geq 2$)}, showing an exponential speedup in the rate of multipartite state distribution for the MP protocols developed. The MP protocols achieve a DR which scales with $O(1)$ for the distance between users, whereas for the SP protocols $\text{DR} \sim O(p^{|R|})$ for a routing solution of size $|R|$. As $|R|$ will depend on the specific topology and location of users, we use $|R| \geq |S|$ to show the speedup is still exponential for the worst-case when $|R|=|S|$. These results are valid when entanglement links are generated with $p$ above the percolation threshold ($p_{c}=0.5$ for grid lattices). For $p<p_{c}$, the DR achieved by the MP protocols decreases with the distance between users. However, the \hbox{DR-distance} scaling still significantly outperforms that of the SP protocol.\\

These distance-independent results were obtained assuming that entanglement links generated with ideal fidelities and subsequent LOCC operations are also error-free. In NISQ-era networks, these assumptions are not generally valid. For the proposed protocols, the fidelity of the distributed GHZ state will depend on the fidelities of the entanglement links used to generate the state \cite{de_Bone_2020} as well as the local operations performed \cite{nain2020analysis}. This means that a true distance-independent distribution of GHZ states will not generally be feasible for NISQ-era networks, where local operations add error. However, we expect the MP protocols to scale with distance significantly better than SP protocols operating in the same scenarios. This is because the MP protocols have a much higher probability of successful routing, independent of the effect of noisy local operations.

\subsection{Multipartite state distribution rate with varied users}
\begin{figure}[htbp]
    \centering
    \includegraphics[width=\linewidth]{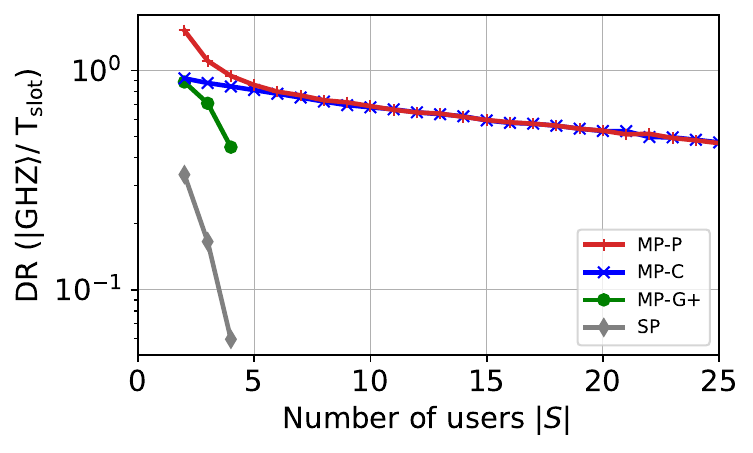}
    \caption{Distribution rate (DR) of $\ket{\text{GHZ}_N}$ states against increasing number of randomly located users, with $N=|S|$ ($M=6, p = 0.75, |S| \in [3,25], Q_c=1$).}
    \label{fig:3}
\end{figure}

The protocols were tested to assess the effect of the number of users on the distribution rate. The number of users was varied between 3 and 25 in a $6\times6$ grid network. The users in $S$ were randomly selected from the set of nodes $V$. As in Section \ref{sec:DRp}, each datapoint represents the average of 500 different sets of users, which all contain the same number of users. \hbox{Fig. \ref{fig:3}} shows the DR achieved by the protocols in this scenario. Similar results were obtained in grid topologies of other sizes. Whereas the standard SP protocol exhibits an exponential decrease in DR with the number of users, the MP-P and MP-C protocols scaled significantly better. Further, the size of the GHZ state generated by the MP-G+ and SP protocols was limited by the number of quantum memories available at the centre node, as we assume one quantum memory per physical edge. In grid networks, these protocols can therefore service up to $|S| = 4$ users, with $|S| = 5$ only feasible if the centre node is also a user.\\ 

As the MP-P and MP-C protocols do not require a central node, they can freely scale with the number of users. Further, in contrast to SP routing, they exhibit a much smaller penalty to the DR for each additional user. For the \hbox{MP-C} protocol, a $\ket{\text{GHZ}_{25}}$ state was generated with an $\text{DR} \approx 0.5 \;$ 
This result is only valid above the critical probability $p>p_{c}$ for percolation, with distributing entanglement between large numbers of users being more challenging otherwise. A further result was that the benefit of MP-P was found to be more significant for fewer users, with a minimal benefit for using the MP-P protocol over MP-C beyond five users in the grid topology due to the absence of multiple disjoint Steiner trees in $G'$. \\

\subsection{Quantum memory decoherence effect on distribution rate} \label{sec:mem}

We have demonstrated that our developed multipath protocols can achieve an exponential speedup, compared to the SP protocol, for sharing multipartite states between multiple users. However, for maximum benefit, entanglement links must be generated with a probability above the critical probability of percolation for the given topology, which is currently infeasible for any realistic quantum network \cite{yu2020entanglement}. In this section, we further demonstrate how the benefits of multipath routing can be observed, below the percolation threshold. This is achievable when the nodes are equipped with quantum memories able to store qubits for multiple timeslots.\\

Previous results consider only entanglement link generation for single independent timeslots ($Q_c=1$) which is a common assumption in the literature \cite{pant2019routing}. However, this limits the possible functionality of the protocols. Patil \textit{et al.} \cite{patil_plus} consider a network model in which entanglement links are attempted for multiple timeslots, but also requires nodes be equipped with an additional quantum memory per edge per timeslot. 
Instead, we consider a single quantum memory per edge that can store an entanglement link over multiple timeslots.

\begin{figure}[thbp]%
    \centering
    \includegraphics[width=\linewidth]{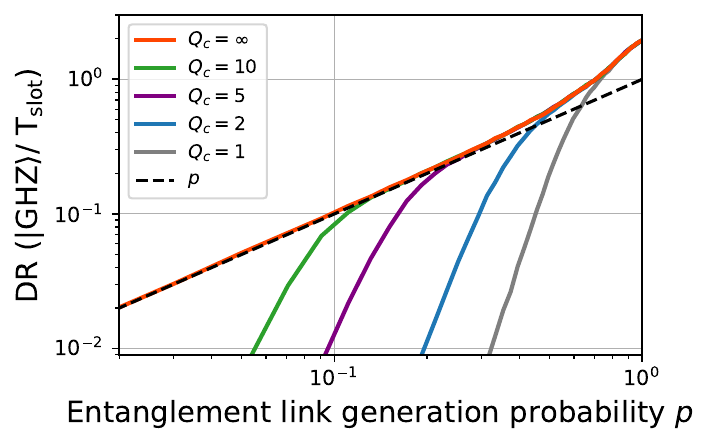}
    \caption{Distribution rate of $\ket{\text{GHZ}_4}$ states by the MP-P protocol against entanglement link generation probability $p$ for 4 randomly located users with selected values of $Q_c$. As a reference, the line $\text{DR} = p$ is also plotted ($M=6, |S|=4$), $p \in [0,1], Q_c \in \{1,2,5,10,\infty\})$.}
    \label{fig:loglog}
\end{figure}

To study the impact of quantum memory decoherence on the distribution rate, we simulated the MP-P protocol for networks with varied quantum memory decoherence times. For each decoherence time $Q_c$ analysed, the value of $Q_c$ was equal for all quantum memories in the network. Results in \hbox{Fig. \ref{fig:loglog}} show that networks with better quantum memories (i.e. higher $Q_c$) achieve a higher DR. However, we observe that increasing $Q_c$ does not improve the DR when the protocol is already generating GHZ states at a DR greater than $p$ (dashed black line). As this condition requires the GHZ state to be generated in fewer than $Q_c$ timeslots, the protocols terminate before higher $Q_c$ can influence the state of $G'$. \hbox{Fig. \ref{fig:loglog}} also shows the protocols achieve a DR which improves rapidly with increasing $p$, until approaching $DR \approx p$ where DR growth decreases. The transition point between these two regimes occurs at approximately $p=1/(Q_c+1)$. \\

This transition can be explained by considering the probability of an entanglement link being present in a specific instance of $G'$, $P(\omega(e)=1)$. When $Q_c=1$, this probability is equal to the entanglement link generation probability $p_e$. However, when entanglement links can be stored for multiple timeslots ($Q_c>1)$, the value of $P(\omega(e)=1)$ also depends on the value of $Q_c$. We derive $P(\omega(e)=1)$ as:

\begin{equation}
P(\omega(e)=1) = \frac{|\{\Omega_T|\omega(e)=1\}|}{|\{\Omega_T|\omega(e)=1\}|+|\{\Omega_T|\omega(e)=0\}|}    
\end{equation}

where $|\{\Omega_T|\omega(e)=x\}|$ is the number of timeslots for which $\omega(e)=x$, with $x  \in \{0,1\}$. For $n$ attempts at entanglement link generation, we expect $n p_e$ to succeed and $n (1-p_e)$ to fail. Each successfully generated entanglement link can remain present over the edge $e$ for $Q_c$ timeslots. Thus, $|\{\Omega_T|\omega(e)=1\}|$, is equal to $n p_e \times Q_c$. As each failed attempt lasts a single timeslot then $|\{\Omega_T|\omega(e)=0\}|$ equals $n (1-p_e)$. Therefore, for a sufficiently large number of attempts, $P(\omega(e)=1)$ is given in (\ref{eq:10}).\\

\begin{equation} \label{eq:10}
    P(\omega(e)=1) = \frac{p_e Q_c}{p_e Q_c + (1-p_e)} 
\end{equation}

In topologies with homogeneous $P(\omega(e)=1)$, percolation can be achieved when $P(\omega(e)=1)>p_c$ \hbox{\cite{bollobas}}. Thus, by making the right side of (\ref{eq:10}) equal to 0.5 we get $p=1/(Q_c+1)$, which explains the observed transition. In Fig. \ref{fig:loglog} we see all protocols achieve similar distribution rates for values of $p$ and $Q_c$, such that $P(\omega(e)=1)>p_c$. Above this critical probability, we expect a single GCC in $G'$, and therefore a high distribution rate of GHZ states is possible. There is a reduced growth in the DR when a GCC exists as there is a high probability all users are already connected. The MP protocols can achieve higher distribution rates over networks where entanglement links can be stored for multiple timeslots, even when the generation probability of the entanglement links is below $p_c$. This shows how improved quantum memories can increase the distribution rates of the developed multipath routing protocols.\\

\begin{figure}[htbp]%
    \centering
    \includegraphics[width=\linewidth]{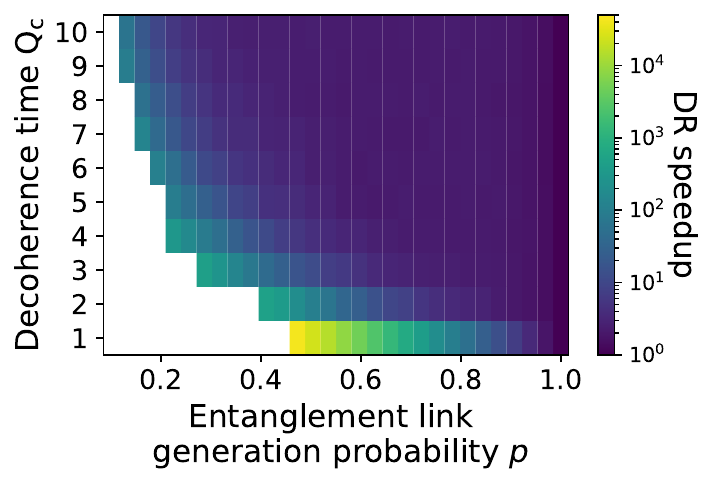}
    \caption{Distribution rate speedup of the MP-P protocol over the SP protocol when distributing $\ket{\text{GHZ}_4}$ states between the four corner nodes of a $6\times6$ grid topology, for a parameter sweep of $p$ and $Q_c$ ($M=6, |S|=4, p \in [0, 1], Q_c \in \{1,2,...10\}$)}
    \label{fig:heat}
\end{figure}

To quantify the relative performance of the MP-P and SP protocol, a wider parameter sweep of $p$ and $Q_c$ was performed in a $6 \times 6$ grid topology with four users located in the corner nodes of the network. \hbox{Fig. \ref{fig:heat}} shows the speedup of the MP-P protocol compared to the SP protocol. The white area shows data points where more than 5\% of simulation runs failed to generate a GHZ state.\\

A speedup in DR was observed for all values of $Q_c \geq 1$ and $0 < p \leq 1$. The largest DR speedup was observed for $p=0.47$, $Q_c=1$, with a $4\times 10^4$ improvement. 
Similar speedups, of different magnitudes, were observed for different-sized grid topologies and for randomly located users. The high DR speedup for these network conditions suggests that the proposed protocols will be useful for NISQ-era networks, where networks consist of devices with short decoherence times and low distribution rates. \hbox{Fig. \ref{fig:heat}} also shows that with larger $Q_c$, the DR speedup becomes significant at lower values of $p$. This shift occurs due to the effect of $Q_c$ on $P(\omega(e)=1)$. While both the MP-P and SP protocols achieve a higher DR with higher $Q_c$, the magnitude of the speedup is reduced mainly due to the relative improvement of the SP protocol. As was shown in Fig. \ref{fig:loglog}, the MP-P protocol can achieve higher DR at low $p$, by increasing $Q_c$. These results demonstrate that the proposed multipath protocols improve the multipartite state distribution rate. 

\section{Performance evaluation: mesh topologies} \label{sec:top}

The protocols developed can be applied to any topology, but the achievable distribution rates might differ depending on the topology. To investigate this we simulated the best performing protocol, MP-P, on topologies taken from real-world optical networks, which are described in Table \ref{tab:1} \cite{baroni1998routing}. The edge lengths of the topologies used were scaled down by a factor of 100, to more closely match the size of current experimental entanglement distribution setups \cite{yu2020entanglement}. The entanglement link generation probability $p_e$ was calculated for each edge using (\ref{eq:1}). To parameterise $p_e$ we vary $p_{\text{op}}$, with $p_{\text{loss}}$ a fixed function of edge length. Further, for the mesh topologies, we only consider networks with $Q_c=1$. The legend identifying the topologies in \hbox{Fig. \ref{fig:X}} also applies to later figures in this section. Performance was evaluated with varied entanglement link generation probabilities and numbers of users.

\begin{table}[htbp]
		\caption{Optical networks topologies with network edge lengths scaled down by a factor of 100, and the $6\times6$ grid topology. \cite{baroni1998routing}}
		\label{tab:1}
		\centering
		     \begin{tabular}{|p{1.4cm}|p{0.9cm}|p{0.9cm}|p{1.8cm}|p{1.2cm}|}
         \hline
        \textbf{Network Name} & \textbf{Number of nodes} & \textbf{Number of edges} & \textbf{Average edge length (km)} & \textbf{Average nodal degree}\\
         \hline
        ARPA & 20 & 31 & 6.09 & 3.1 \\
        EON & 20 & 39 & 7.24 & 3.9\\
        Eurocore & 11 & 25 & 4.26 & 4.55\\
        NSFnet & 14 & 21 & 5.09 & 3.0\\
        UKnet & 21 & 39 & 1.38 & 3.71\\
        USnet & 46 & 76 & 4.34 & 3.3\\
        Grid-6 \hbox{(Sec. \ref{sec:DRp})} & 36 & 60 & 1.0 & 3.33 \\
         \hline
    \end{tabular}
\end{table}

\subsection{distribution rate speedup in mesh topologies} 

\begin{figure}[th]
    \centering
    \includegraphics[width=0.49\textwidth]{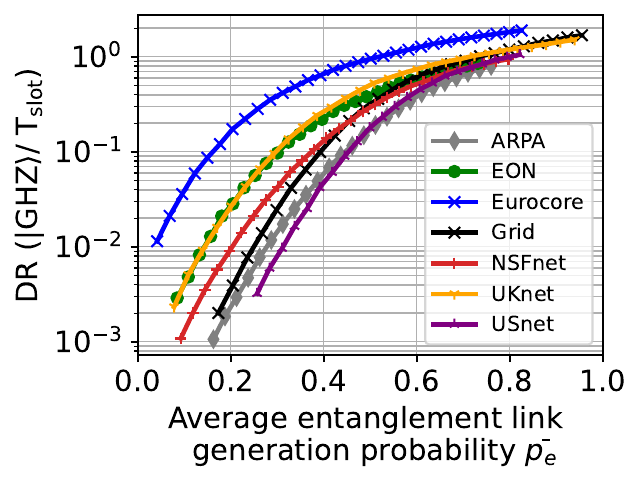}
    \caption{Distribution rate (DR) of  $\ket{\text{GHZ}_5}$ states by the MP-P protocol against average entanglement link generation probability $\bar{p_e}$ for 5 users randomly located in each topology. $\bar{p_e}$ was varied by sweeping the parameter $p_{\text{op}}$. The legend used also applies to later figures in Section \ref{sec:top} \hbox{($|S|=5, p_{op} \in [0,1], Q_c=1$)}.} \label{fig:X}
\end{figure}

We plot the DR against average edge probabilities $\bar{p_e}$, to allow for easier comparison among topologies with a variety of edge lengths, and therefore values of $p_e$. We used randomly located users with $|S|=5$ and $Q_c=1$. \hbox{Fig. \ref{fig:X}} shows that the DR followed a similar trend as seen in the grid topologies with uniform $p$. However, the absolute DR achieved by the protocols varied among topologies, especially at low $\bar{p_e}$. Further, the topologies in which the highest DR was achieved (Eurocore, EON and UKNet) had a wide range of average edge lengths among them. This suggested that multiple factors, such as nodal degree, edge length, and network size, all affect the achievable DR.\\ 

\begin{figure}[th]
    \centering
    \includegraphics[width=0.49\textwidth]{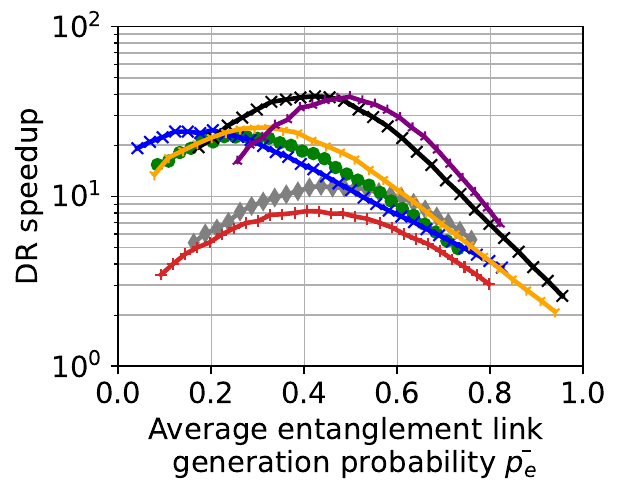}
    \caption{Distribution rate speedup of the MP-P protocol over the SP protocol against $\bar{p_e}$ when distributing $\ket{\text{GHZ}_5}$ states between 5 randomly located users \hbox{($|S|=5, p_{op} \in [0,1], Q_c=1$)}.} \label{fig:22}%
\end{figure}

The speedup of the MP-P protocol over the SP protocol on the varied topologies is shown in \hbox{Fig. \ref{fig:22}}. For the mesh topologies we compare the DR against the tree-variant of the SP routing \cite{meignant2019distributing}. In this approach, routing is performed along the Steiner tree in $G$. We use this more general protocol in the mesh topologies as it is more robust to networks with varied edge degrees. Hence the location of a valid centre node will not affect the distribution rate achieved by the SP protocol.\\

For all topologies, a significant speedup occurred for all values of $\bar{p_e}$, with a maximum at an intermediate value of $\bar{p_e}$. Depending on topology, the maximum speedup was observed between $0.28<\bar{p_e}<0.53$. For the Grid-6 network, the maximum speedup occurred at $\bar{p_e}=0.52$, close to the percolation threshold for this topology. The speedup achieved by the MP-P protocol was found to be reduced for both high and low $\bar{p_e}$. For high $\bar{p_e}$ the SP protocol was sufficient to obtain a high DR, hence the relative speedup achievable for the MP-P protocol decreased. 
Similarly, for very low $\bar{p_e}$, we suggest that routing will predominately succeed along the minimum distance tree in $G'$, as longer trees will exist with a significantly lower likelihood. This also reduces the benefit of the proposed multipath protocols. However, the results suggest a significant DR improvement will still be achieved for low $\bar{p_e}$. At intermediate values of $\bar{p_e}, $ there was sufficient redundancy in the edge distribution of the subgraph $G'$, such that routing succeeded with a high rate for multipath protocol. In contrast, the SP protocol has a much reduced DR as a single entanglement link failure along the route prevents a GHZ state from being generated. The magnitude of the speedup was found to be lower compared to that seen in \hbox{Fig. \ref{fig:heat}}. This is primarily due to the choice of users rather than a feature of the topologies. By using randomly located users, instead of users in distantly located nodes, the average distance between users was lower and hence the magnitude of the speedup was reduced.

\subsection{Scaling with users in mesh topologies}

\begin{figure*}[ht]%
    \centering
    \subfloat[\centering]
    {{\includegraphics[width=0.3\textwidth]{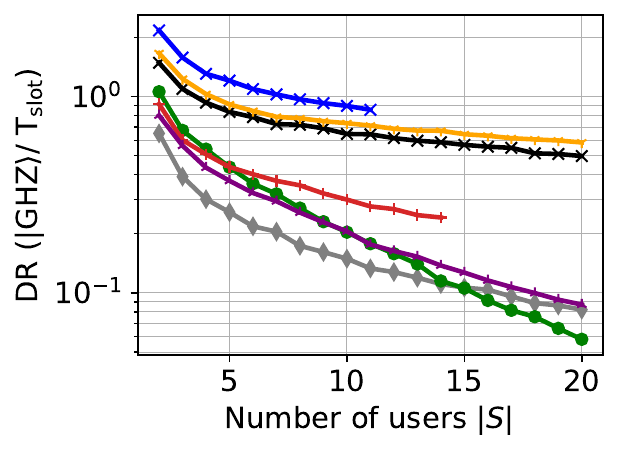}\label{fig:31}}}
    \qquad
    \subfloat[\centering]{{\includegraphics[width=0.3\textwidth]{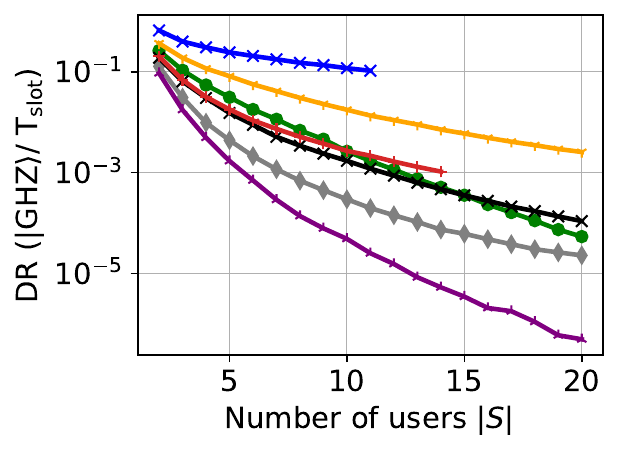}\label{fig:32}}}
    \qquad
    \subfloat[\centering]{{\includegraphics[width=0.3\textwidth]{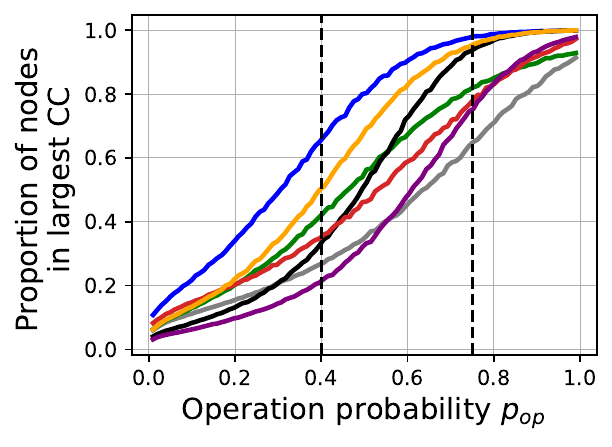}\label{fig:33}}}
    \qquad
    \caption{(a) Distribution rate of $\ket{\text{GHZ}_N}$ states, N=|S|, for the MP-P protocol with number of users $|S|$ and $p_{\text{op}}=0.75$ (b) Distribution rate of $\ket{\text{GHZ}_N}$ states, N=|S|, for the MP-P protocol with number of users $|S|$ and $p_{\text{op}}=0.4$ (c) Proportion of nodes that are part of the giant connected component (GCC) in $G'$ \hbox{($Q_c =1$)}.} 
    \label{fig:30}%
\end{figure*}

The MP-P protocol was further studied on the mesh topologies to assess the impact of the number of users on the DR.
Performance was evaluated for two values of operational probability $p_{\text{op}}$ as shown in \hbox{Fig. \ref{fig:30}}, a) $p_{\text{op}}=0.75$ and b) $p_{\text{op}}=0.4$. The results show that the DR decreases for additional users, as seen in \hbox{Fig. \ref{fig:3}} for the grid topology. Additionally, there is a significant variation in the DR scaling behaviour between topologies. Certain topologies such as the Eurocore, UKnet and Grid-6 networks achieve high distribution rates even for a large number of users. This was thought to be primarily due to their higher nodal degree. For the multipath protocols, being able to utilise many possible paths means that a high nodal degree improves DR.\\

However, the ordering of DR achieved by the protocols in different topologies was not consistent for all values of $p_{\text{op}}$. For example, \hbox{Fig. \Ref{fig:31}} shows the grid topology (black line) performed better at $p_{\text{op}}=0.75$ relative to the other topologies than at $p_{\text{op}}=0.4$ \hbox{(Fig. \ref{fig:32})}. This behaviour might be explained by considering the size of the largest connected component of $G'$, which for these networks will be a function of $p_{\text{op}}$ and topology \cite{stauffer2018introduction,newman2001random}. \hbox{Fig. \ref{fig:33}} shows the proportion of network nodes belonging to the largest connected component with varied $p_{\text{op}}$. These mesh topologies do not have defined percolation thresholds, with the proportion of nodes in the largest connected component following a continuous distribution with $p_\text{op}$. 
However, the size of the largest connected component at $p_{\text{op}}=0.4$ and $p_{\text{op}}=0.75$ correlates with variation in relative performance observed between the topologies. In Section \ref{sec:analytical}, we relate the relationship between the distribution rate DR and the connected component size analytically.

\section{Analytical Distribution Rate} \label{sec:analytical}
The Monte Carlo simulation model allows for varied protocols and network models to be simulated. However, this approach is computationally intensive when simulating low DR values. Therefore, analytical expressions for the DR were reviewed in Section \ref{sec:ub} and proposed in Section \ref{sec:approx}. The developed analytical approach allowed fast comparison of network parameters, such as topology, on the protocols.\\

\subsection{Analytical approximations} \label{sec:ub}
We define the upper bound of the DR as the maximum number of GHZ states that can be distributed per timeslot. From the graph definition of the network, this upper bound equals the number of edge-disjoint Steiner trees connecting the users in $G$. Similarly, for a given timeslot, the number of GHZ states that can be distributed is the number of such trees in $G'$. Finding the number of edge-disjoint trees in a graph is the Steiner tree packing problem \cite{packing}. However, due to the hardness of solving this problem directly, an equivalent upper bound was instead considered. This equivalent is the minimum cut of edges in $G$ to separate a user from the set of other users. For the grid topology, this min-cut is the nodal degree of a user, giving an upper bound of $\text{DR} \leq 4 P(\omega(e)=1)$. The protocols developed were not found to approach this upper bound for any values of $p$. This suggests that routing multiple GHZ states in $G'$ is challenging, even for networks where entanglement links are present with a probability above the critical threshold for percolation. \\

Given the performance of the previous bound, a separate approach considers estimating DR by calculating the probability of all users $S$ being in the same connected $C$ component, $P(S \in C)$. For graphs with probabilistic edges, finding this probability is known as the k-terminal reliability problem \cite{reed2019efficient,zenklusen2008combinatorial}. In general topologies, this problem is NP-hard, but if percolation can be assumed then simplified models can be utilised \cite{newman2018networks}. For infinite lattices in percolation, the term $\theta(p)$ gives the probability of a node being part of the GCC \cite{grimmett2013percolation}. The probability of all users being in this GCC is given by $P(S \in GCC) = \theta(p)^{|S|}$. However, in the finite topologies considered this expression is not accurate for values of $p<p_c$. Therefore, it cannot be used for low computational complexity estimation of DR for the scenarios considered in this paper. 

\subsection{improved approximation} \label{sec:approx}

For small graphs, the term $\theta(p)^{|S|}$ can not be used to find $P(S \in GCC)$. We can instead substitute $\theta(p)$ with $|C|/|V|$, where $C$ is the largest connected component in $G'$. However, this approach also assumes that each event $(s \in C)$, $\forall s \in S$ is independent, which is not the case when $|S| \approx |C|$, such as below percolation.\\

Thus, to get a better approximation for the probability $P(S \in C)$, we modelled this as a \textit{Hypergeometric discrete probability distribution} \cite{ross2014introduction}. The Hypergeometric distribution describes the probability of k successes in n draws \textbf{without replacement} from an X-sized population. Making $k=|S|$, $n=|C|$ and $X=|V|$, this is equivalent to the probability of obtaining the nodes of $S$, after randomly selecting $|C|$ nodes from $V$ without replacement. Therefore, the probability of all $|S|$ nodes of $S$ being in a connected component $C$ is given by:

\begin{equation} \label{eq:x1}
P(S \in C) \approx M(V,S,C) = \frac{\binom{|V|-|S|}{|C|-|S|}}{\binom{|V|}{|C|}} 
\end{equation}

The term  $\binom{|V|}{|C|}$ gives the number of combinations of unique connected components of size $|C|$ that can be found from the nodes $V$. Similarly, $\binom{|V|-|S|}{|C|-|S|}$ is the number of ways a set $S$ can be arranged in $C$. Combined, these expressions give the proportion of events for which a random connected component of size $|C|$ includes all users $S$. \\

For the MP-C and MP-P protocols, a valid routing solution only requires that all users are in the same connected component of $G'$. Below the percolation threshold, no GCC exists and therefore the size of the largest connected component $C$ will be some probability distribution that is a function of $p$ and the network topology. The probability of the largest connected component in $G'$ having $i$ nodes ($|C_i|$) is given by \hbox{$\alpha_i = P(|C_i|=i)$} where $\sum^{|V|}_{i=1} \alpha_i = 1$. We found the values of $\alpha_i$ by numerically simulating $G'$. The DR can then be estimated using a weighted sum of (\ref{eq:x1}) with $\alpha_i$ as the weights:

\begin{equation} \label{eq:x2}
\text{DR} \approx \sum_{i = 1}^{|V|} \alpha_i \times M(V,S,C_i)
\end{equation}

\begin{figure}[htbp]%
    \centering
    \includegraphics[width=\linewidth]{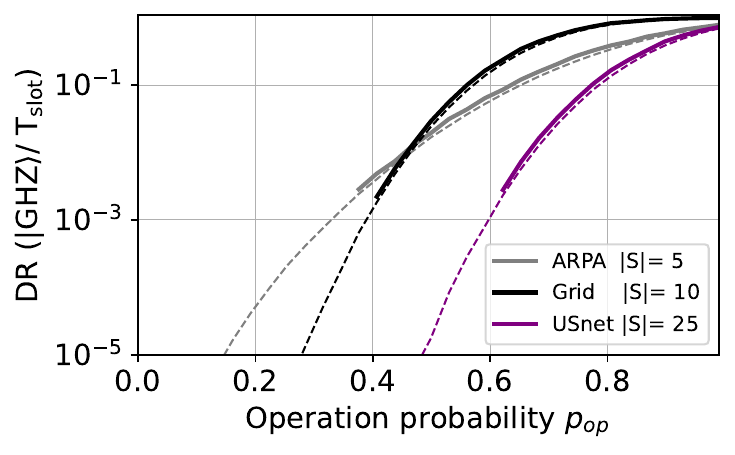}
    \caption{Distribution rate of $\ket{\text{GHZ}_N}$ states, N=|S|, of the MP-C protocol versus operational probability $p_{\text{op}}$. The distribution rate of the \hbox{MP-C} protocol was calculated using both the Monte Carlo simulation (solid line) and using (\ref{eq:x2}) (dashed). The values were found for a subset of topologies and number $|S|$ of randomly located users ($Q_c=1$).}
    \label{fig:simulated}
\end{figure}

Fig. \ref{fig:simulated} shows the closeness of fit between the DR calculated by Monte Carlo simulation of the MP-C protocol, and from (\ref{eq:x2}). This means this analytical expression can be used to calculate the DR where it is not computationally efficient to use a Monte Carlo simulation. More exact methods may be used that work for networks that operate strictly above the percolation threshold \cite{newman2001random,newman2018networks,kitsak2018stability}.\\

\section{Conclusions} \label{sec:conclusions}
We propose three protocols, MP-G+, MP-C, and MP-P, for the distribution of shared multipartite states across a quantum network. The protocols were designed to use multipath routing, in order to improve the multipartite distribution rate. These protocols were simulated on quantum networks, modelled to consider probabilistic Bell pair distribution and qubit decoherence, such as would be observed in a network of NISQ devices.\\

The effect of network topology on the proposed protocols was assessed by simulating the protocols on topologies taken from real optical networks. The results show that the achieved distribution rate varied between topologies, with features such as network size and average nodal degree found to have an effect. An analytical approximation using the probability distribution of successfully distributed Bell pairs was found to quantify these effects.\\

Results show that the proposed protocols all achieved an exponential speedup in the rate of multipartite state distribution. This speedup was observed with the distance between users when compared to protocols using single path routing. The observed speedup increased for many users, and when Bell pairs were distributed with a probability close to the percolation threshold for the given network topology. Of the protocols developed, the MP-P protocol achieved the highest rate of multipartite state distribution and when simulated achieved a speedup of up to four orders of magnitude. The use of quantum memories with improved decoherence times was considered. While improved quantum memories improved the multipartite distribution rate for all protocols considered, the speedup of the multipath protocols was most significant for short decoherence times. As the multipath protocols provide speedup for intermediate values of entanglement generation and short decoherence times, this research will have possible applications for NISQ quantum networks. \\

\bibliographystyle{IEEEtran}
\bibliography{IEEEabrv,bib}
\EOD

\end{document}